\documentclass[a4paper]{article}
\usepackage{amsfonts}
\usepackage{amssymb}
\usepackage[margin=2cm]{geometry}
\usepackage[dvips]{graphicx}

\usepackage[T1]{fontenc}
\usepackage[latin2]{inputenc}

\usepackage{amsmath}
\usepackage{bbm}

\begin{document}


\title{The macroscopic qubit delusion}

\author{Robert Alicki \\ 
  {\small
Institute of Theoretical Physics and Astrophysics, University
of Gda\'nsk,  Wita Stwosza 57, PL 80-952 Gda\'nsk, Poland}\\
}

\date{\today}
\maketitle

\begin{abstract}
There exists a large number of experimental and theoretical results supporting the picture of "macroscopic qubits" implemented, for instance, by
Rydberg atoms,  Josephson junctions or Bose-Einstein condensates - the systems which should rather emerge in localized semiclassical states. In this note it is shown how, under realistic conditions, the false qubit interpretation can be consistent with the restricted set of experimental data collected for semiclassical systems. The recent experiments displaying semiclassical character of Bose-Einstein condensates and  possible quantumness tests for a single system are briefly invoked also.
\end{abstract}

In the last decade remarkable experiments were performed involving measurements and manipulations of states for  single physical systems which were identified with  simple quantum mechanical systems described by  low dimensional Hilbert spaces. These systems can be divided into two categories. The first one consists of those which obviously belong to the quantum domain like atoms or ions at the lowest energy levels, single-photon polarization, particle's spins , electrons in quantum dots, or single mod of radiation
at low numbers of photons. This note is entirely devoted to the other class which contains either small systems excited to high quantum numbers or many-body systems, in both cases  expected to be  rather observed
in well-localized semiclassical states which seem to be the only relatively stable with respect to  external noise. The examples are: Rydberg atoms at circular states used in quantum-optical experiments \cite{H},
mesoscopic Josephson junctions \cite{W} and Bose-Einstein condensate in a double-well potential \cite{Chen}. For simplicity we shall concentrate ourselves on the cases where  phenomenology of such systems is described in terms of two-level quantum systems (qubits) with suggested applications to quantum information processing. 
\par
First, the mechanism will be outlined which can lead to a consistent description of experimental data in terms of a qubit model despite the semiclassical character of the real system. Then, the discussion of particular examples follows.
\par
\emph{Spin-$j$ model}

The canonical model of the discussed systems is a  spin-$j$  (with half-integer $j >> 1$) defined by angular momentum operators $J_k , k=1,2,3$
acting on the $(2j+1)$-dimensional Hilbert space with the basis $|j,m\rangle , m = -j, -j+1,...,j$. 
The typical  Hamiltonian can be approximated by the following second-order polynomial in $J_k$
\begin{equation}
H = \Omega J_3^2 + \Delta J_3 + \Gamma J_1
\label{Ham}
\end{equation}
with real parameters $\Omega, \Delta , \Gamma$ and the system is controlled by the time-dependent Hamiltonian  of the form
\begin{equation}
H_c(t) = \sum_{k=1}^3 h_k(t)J_k . 
\label{gateham}
\end{equation}
with real control fields $h_k(t)$.
\par
We compare now two pictures:

I) \emph{Macroscopic qubit}  The two states $|j,1/2\rangle$ and $|j,-1/2\rangle$ are well-separated from the others, their superpositions can be prepared, they are relatively stable with respect to the environmental noise and approximatively invariant under the dynamics.

II) \emph{Semiclassical system}  The experimentally accessible, relatively stable states are semiclassical, localized ones with fluctuations $\langle (\Delta J_k)^2\rangle = \mathcal{O}(j)$
which approximatively follow classical trajectories. In addition the accessible states have supports on the subspace ${\cal H}_{\delta m}$ spanned by the basis vectors  $|j, m\rangle$ with  $|m|\leq \delta m \simeq \mathcal{O}(\sqrt{j})$.

\par 
The assumptions behind the first picture are very difficult to justify both mathematically and physically but provide a simple model which explains quite well the experimental data and therefore is rather convincing. For this reason, the point of view I) is adopted in most of the papers (for notable exceptions see \cite{G1,G2,RA}). On the other hand there are numerous theoretical arguments supporting the stability of semiclassical states (e.g. \cite{RA, Ben1, Ben2}), but the explanation is needed how to justify within the second picture the agreement of experimental data with the first one.
\par
Phenomenology of the discussed systems involves always a measurement of a fixed \emph{unsharp observable} denoted by  $\mathcal{S}$ with two outcomes $\pm 1$ and the corresponding positive operator-valued measure $\{S_+ , S_- \geq 0,S_+ + S_- = I\}$. As $\mathcal{S}\equiv \mathcal{S}^3$ one can choose the unsharp sign of the spin component $J_3$ 
\begin{equation}
S^3_{\pm} =  \frac{1}{2}(I\pm F(J_3))
\label{S3}
\end{equation}
determined by the \emph{sensitivity function} $F(-x)= -F(x)$ which monotonically grows from the value $-1$ to the value $1$.

 Another ingredient consists of \emph{quantum gates}
-the unitary maps describing evolution of the system governed by the total Hamiltonian between initial and final time
\begin{equation}
U(t_{in},t_{fin}; h_k) = \mathbf{T}\exp -i\int_{t_{in}}^{t_{fin}} (H + H_c(t))dt.
\label{gate}
\end{equation}
In the semiclassical regime ($j >> 1$) operators $J_k$ and the Hamiltonian $H + H_c(t)$ possess classical limit and the evolution of the semiclassical
state can be approximated by the motion of the unit vector ${\bf n} \simeq \langle \bf{J}\rangle /j$ satisfying the classical equation of motion in the form
\begin{equation}
\frac{d{\bf n}}{dt} = \bigl(j\Omega ({\bf n}{\bf e}_3) + \Delta {\bf e}_3+\Gamma {\bf e}_1 + {\bf h}(t)\bigr)\times {\bf n}.
\label{gate2}
\end{equation}
By a proper tuning of the parameters $t_{in},t_{fin} h_k(t)$ one can produce a gate $U_1$ approximatively describing the effect of rotation which transform $\bf{e}_3$ into $\bf{e}_1$, leaving $\bf{e}_2$ invariant (analogically the gate $U_2$). One puts $U_3=I$. This allows to define  three unsharp observables ${\cal S}^k$ by combining gates with the measurement of  ${\cal S}^3$
\begin{equation}
S_{\pm}^k = U_k^{\dagger} S_{\pm}^3 U_k ,\quad k=1,2,3 .
\label{S12}
\end{equation}
For a given state $\rho$ of the spin-$j$ one can perform a \emph{restricted tomography} by  measuring the mean values of the observables $\mathcal{S}_k$ for $k=1,2,3$, called Stokes parameters
\begin{equation}
s_k = \langle \mathcal{S}_k\rangle =\mathrm{Tr}(\rho S_+^k)-\mathrm{Tr}(\rho S_-^k) =\mathrm{Tr}(U_k\rho U_k^{\dagger} F(J_3))
\label{fuzzy}
\end{equation}
Applying the expansion
\begin{equation}
 F(J_3))= F'(0)J_3 + \frac{1}{6}F'''(0) J_3^3 +\cdots
\label{exp}
\end{equation}
one can compute
\begin{equation}
s_k \simeq F'(0)\mathrm{Tr}(U_k\rho U_k^{\dagger} J_3).
\label{fuzzy1}
\end{equation}
The Stokes parameters satisfy the inequality
\begin{equation}
s_1^2 +s_2^2 +s_3^2 \leq \min\{ 3, 3(F'(0))^2 \delta m^2\}
\label{ine2}
\end{equation}
where the first bound follows from the definition (\ref{fuzzy}) while the second one is based on the approximation (\ref{fuzzy1}) and the initial assumption II). 
One can define the following qubit's density matrix
\begin{equation}
\rho_{q} = \frac{1}{2}( I +\vec{s}\cdot \vec{\sigma}).
\label{qubit}
\end{equation}
which makes sense under the normalization condition 
\begin{equation}
s_1^2 +s_2^2 +s_3^2 \leq 1.
\label{ine3}
\end{equation}
Although the condition  (\ref{ine3}) does not follow immediately from (\ref{ine2}) there are several reasons why (\ref{ine3}) is satisfied under realistic conditions.
First of all the raw experimental data are proceeded using the different types of normalization, maximum likehood techniques, proper fitting, e.g. "including an offset accounting for residual noise", etc., which can enforce the condition (\ref{ine3}). The other factor is unprecise
preparation of the initial state which reduces the values of $|s_k|$.
The dynamics of $s_k$ can be derived using the semiclassical equation (\ref{gate2}) and (\ref{fuzzy1}) to obtain a  nonlinear evolution equation of the form (\ref{gate2})
with ${\bf n}$ replaced by ${\bf s}$ and $\Omega$ by  $\Omega/F'(0)$. 
The linearized version of such evolution can be misinterpreted as the Bloch equation for the qubit density matrix (\ref{qubit}). The nonlinear term combined with the quantum fluctuations of ${\bf s}$ of the order ${\cal O}(\sqrt{1/j})$  explain the notorious  large
phase damping in comparison to the energy damping observed for Rydberg atoms and Josephson junctions.

\par
\emph{Rydberg atoms}

Atoms in circular Rydberg states $|n\rangle$ are characterized by a single quantum number $n$
which is assumed to be large ($n > 30$) and the energy 
\begin{equation}
E_n = -\frac{R}{(n-\delta)^2} . 
\label{Ryen}
\end{equation}
Assuming that the experimentally accessible states are superpositions (mixtures) of $|n\rangle$
with $|n-n_0| << n_0$ we can  use the expansion 
\begin{equation}
E_n = -\frac{R}{(n_0-\delta)^2} +\frac{2R}{(n_0-\delta)^3}(n-n_0)-\frac{6R}{(n_0-\delta)^4}(n-n_0)^2 +\dots 
\label{Ryen1}
\end{equation}
to obtain a spin-$j$ ($j= n_0 +1/2$) representations of the atomic Hamiltonian in the form (\ref{Ham}) with
\begin{equation}
\Delta= \frac{2R}{(n_0-\delta)^3},\quad \Omega = -\frac{3\Delta}{(n_0 - \delta)}<<\Delta, \quad \Gamma =0 .
\label{Ryham}
\end{equation}
The controll by means of the external electromagnetic fields  leads to dipole transitions $n\to n\pm 1$ and therefore can be described by the time-dependent Hamiltonians of the form
\begin{equation}
H_c(t) = h_1(t) J_1+ h_2(t) J_2  .
\label{Rycon}
\end{equation}

The measurement technique used in the experiments with Rydberg circular states is based on the selective field ionization which allows to approximatively distinguish the states with $ n\geq n_0 +1$ from the states with $n \leq n_0$. It seems that the unsharp observable (\ref{S3}) is a perfect model of this experimental setting. 
\par

\emph{Superconducting qubits}

As an example of "superconducting qubit" one can take a Cooper pair box which is a circuit consisting of a small superconducting island connected  via Josephson junction to a large superconducting reservoir . Coulomb repulsion between Cooper pairs in a small electrode become important and must be taken into account in the Hamiltonian. The simple Josephson Hamiltonian reads \cite{W}
\begin{equation}
H = E_C\sum_n( n- n_0 -1/2)^2 |n\rangle\langle n|  - {E_J}\sum_n ( |n+1\rangle\langle n| +|n\rangle\langle n+1|)
\label{cpbham}
\end{equation}
where $|n\rangle$ describe the state with $n$ Cooper pairs on the island, $E_C$ determines the magnitude of the Coulomb repulsion, $E_J$ governs the tunneling process,  $n_0 >> 1$ is a number of Cooper pairs on the island at the neutral reference state and the additional "1/2" comes from the standard fine tuning of the system. Under the assumption $n_0 >> 1$ and restricting to the states with $|n-n_0| << n_0$ the Hamiltonian (\ref{cpbham}) can be rewritten in terms of spin variables with $j=n_0 + 1/2$
\begin{equation}
H = E_C J_3^2  - \frac{E_J}{j}J_1
\label{cpbham1}
\end{equation}

The device is controlled by external electromagnetic fields which  are coupled to the net electric charge $Q= 2e J_3 $ and to the electric current $\frac{dQ}{dt}= i[H,Q]\sim J_2$. Hence the control Hamiltonian is given by 
\begin{equation}
H_c(t) = h_3(t)J_3 + h_2(t)J_2.
\label{Rycon1}
\end{equation}
The standard measurement using a single-electron transistor
allows to approximatively determine the sign of the net charge on the island which is exactly the unsharp sign of $J_3$ given by (\ref{S3}).

\par
\emph{Bose-Einstein condensate}

A Bose-Einstein condensate of  $N$ ultracold atoms in a symmetric double-well potential can be described by the two-mode Hubbard Hamiltonian with 
two pairs of annihilation and creation operators $\{a, a^{\dagger}, b, b^{\dagger}\}$. Introducing the fictitious spin components
\begin{equation}
J_1 = \frac{1}{2} (a^{\dagger}b + b^{\dagger}a),\quad  J_2 = \frac{i}{2}(a^{\dagger}b - b^{\dagger}a), \quad
J_3=\frac{1}{2}(a^{\dagger}a - b^{\dagger}b)
\label{BE}
\end{equation}
one can treat the system as a large spin with $j = [N/2] +1/2$ and the Josephson Hamiltonian of the form (\ref{cpbham1}). Again one obtains the same mathematical scheme which leads to the false qubit picture when the unsharp measurement of the sign of the atom number difference is introduced.

Fortunately, in the case of BEC much more precise experimental results exist supporting the semiclassical character of accessible states  \cite{Es}. They show that those states are squeezed spin states with the fluctuations of all spin components of the order $\sqrt{j}$.
As all presented models are  mathematically equivalent this is a strong argument against the macroscopic qubit picture for the previous examples as well. On the other hand if one assumes that the macroscopic qubit picture is correct and such systems could be useful for quantum information processing, then it should be possible to apply quantumness tests like those proposed in \cite{A1, A2} and realized for the case of single photon polarization in \cite{Gen1,Gen2}.

\textbf{ Acknowledgements} This work is supported by the Polish research network LFPPI.

\end{document}